\documentstyle[amssymb,aps,prb]{revtex}

\begin{document}
\title{Positively charged magneto-excitons in a \\ semiconductor
quantum well.} 
\author{C. Riva\cite{clara} and F. M. Peeters\cite{francois}}
\address{Departement\ Natuurkunde, Universiteit Antwerpen (UIA),
B-2610 Antwerpen, Belgium}
\author{K. Varga\cite{kvarga}}
\address{Solid State Division, Oak Ridge National Laboratory, Oak Ridge, Tennessee 37831-3062}
\address{}
\date{\today }
\maketitle

\begin{abstract}
A variational calculation of the lower singlet and triplet states
of positively charged excitons (trions) confined to a single
quantum well and in the presence of a perpendicular magnetic
field is presented. We study the dependence of the energy levels
and of the binding energy on the well width and on the magnetic
field strength. Our results are compared with the available
experimental data and show a good qualitative and quantitative
agreement. A singlet-triplet crossing is found which  for a 200
\AA \ wide GaAs is predicted to occur for B$=15$ T.

\end{abstract}
\twocolumn
\section{Introduction}

In spite of the large number of
theoretical\cite{Lampert,Stebe75,Stebe89,Japan,Whittaker,Stebe2000,Dzyub1,Dzyub2,wojsz,riva2,riva3}
and
experimental\cite{Kheng,Shields95-2,Shields95-1,Govaerts,Finkelstein,Shields,Osborne,Glasberg,manus1,Kim-chapter3,Vanhoucke,Muntenau}
works published in recent years on the subject of charged excitons
in quantum wells, only a very small number of them deals with
positively charged
excitons\cite{Japan,Stebe2000,riva2,Shields95-1,Finkelstein,Osborne,Glasberg}.
Different authors\cite{Japan,Osborne} used the diffusion Monte
Carlo technique to calculate the dependence of the binding energy
of the positively charged exciton on the well width at zero
magnetic field. St{\'e}b{\'e} and Moradi\cite{Stebe2000} reported
low magnetic field results for the positively charged exciton
spin-singlet state, which was calculated using a deterministic
variational technique. To our knowledge, this is the only
theoretical calculation on the magnetic field dependence of the
$X^+$ energy. In the present paper we go beyond the small
magnetic field limit and present results for the whole magnetic
field range not only for the singlet but also for the triplet
state.

 Experimental evidence of the positive
trion ($X^+$) in GaAs/AlGaAs quantum wells was found by different
groups\cite{Shields95-1,Osborne,Finkelstein,Glasberg}. These
results  confirm the existence of both the spin-singlet  state
and the spin-triplet state at non zero magnetic field as was also
found for the $X^-$. The magnetic field dependence of the binding
energy, on the other hand, can be rather different for the $X^+$
as compared to the $X^-$.

We extended our numerical technique, see Ref.~\onlinecite{riva3},
which we used for negatively charged excitons, to describe
positively charged excitons in quantum wells under the influence
of a magnetic field parallel to the quantum well axis. The present
paper is organized as follows. In Sec.~\ref{first-P} we present
the Hamiltonian of the problem. The dependence of the binding
energy on the well width is discussed in Sec.~\ref{first-PP}. In
Sec.~\ref{first-PPP} the magnetic field dependence is
investigated and we discuss the dependence of the average distance
between different pairs of particles in the system as a function
of the magnetic field. We point out the differences and
similarities between the $X^+$ and the $X^-$. The behaviour of the
pair correlation functions is also studied for different magnetic
fields. In Sec.~\ref{second-P} we compare our results with the
experimentally measured transition and binding energies. Finally,
in Sec.~\ref{third-P} we summarize our results and present our
conclusions.

\section{Hamiltonian}
\label{first-P}

In the effective mass approximation  the Hamiltonian describing a
positively charged exciton, i.e. X$^{+}$, in an uniform magnetic
field B is:
\begin{equation}
H=\sum_{i=1}^{3}{\frac{1}{2m_{i}}}({\vec{p}}_{i}-{\frac{e_{i}}{c}}{\vec{A}}
_{i})^{2}+\sum_{i=1}^{3}V({\vec{r}}_{i})+\sum_{i<j}{\frac{e_{i}e_{j}}{%
\varepsilon |{\vec{{r}}}_{i}-{\vec{{r}}}_{j}|},}
\end{equation}
where ${\vec{A}}_{i}={\frac{1}{2}}{\vec{r}}_{i}\times {\vec{B}}$;
$m_{i},e_{i}$ are the masses and charges of the interacting
particles and $\varepsilon$ is the dielectric constant of the
medium the particles are  moving in and is taken the same for the
well and barrier material. The confining potential is
\begin{equation}
V({\mathbf{r}}_i)=0\ \ \ \mathrm{if}\ \ |z|<W/2\ \ \ \mathrm{and}\
\ \ V({\mathbf{r}}_i)=V_{i}\ \ \ \mathrm{if}\ \ \ |z|>W/2,
\end{equation}
where W is the quantum well width, and the reference system is
taken such that the origin of the coordinate system is at the
center of the quantum well. For a GaAs/Al$_{x}$Ga$_{1-x}$As
quantum well the heights of the square well confinement potentials
are $V_{e}=0.57\times (1.155x+0.37x^{2})$ eV for the electrons and
$V_{h}=0.43\times (1.155x+0.37x^{2})$ eV for the hole. If we
consider the case when the magnetic field is applied along the
growth axis of the well, i.e. $ \vec{B}{=(0,0,B)}$, the
Hamiltonian becomes:
\begin{equation}
H=\sum_{i=1}^{N}{\frac{1}{2m_{i}}}\left( -\hbar ^{2}\Delta _{i}+{\frac{%
e_{i}^{2}B^{2}}{4c^{2}}}(x_{i}^{2}+y_{i}^{2})+{\frac{e_{i}\hbar B}{c}}%
l_{zi}\right) +\sum_{i=1}^{N}U({\mathbf{r}}_{i})+\sum_{i<j}{\frac{e_{i}e_{j}}{%
|{\vec{{r}}}_{i}-{\vec{{r}}}_{j}|},} \label{ham-chapter4}
\end{equation}
where $l_{zi}$ is the component along the $z$-axis  of the orbital
momentum of the $i-$th particle. The same considerations made in
Ref.~\onlinecite{riva3} for the $X^-$ hold also in this case and
the same functional form for the trial wave function which
consists of a linear combination of deformed correlated wave
functions is used to solve the Hamiltonian. However, here the
variational parameters $A_{ij}$ which enter in the definition of
the deformed gaussian functions can assume also negative values.
For our numerical calculation we focus on a
GaAs/Al$_x$Ga$_{1-x}$As quantum well with $x=0.3$. The parameters
used are $\varepsilon =12.58$, $m_{e}=0.067$ $m_{0}$ and $
m_h=0.34m_0$, which result in $2R_{y }=e^2/\varepsilon a_B=11.58$
meV and $a_{B }=\varepsilon \hbar^2/e^2m_e=99.3$~{\AA}.

\section{Zero magnetic field trion energy}
\label{first-PP}

First, we study the binding energy of the positively charged
exciton which is defined as
\begin{equation}
E_B(X^+,B)=E(X)+E_h(W,B)-E(X^+),
\end{equation}
where $E(X)$ and $E(X^+)$ are respectively, the total energy of an
exciton and of a positively charged exciton in the quantum well
and $E_h(W,B)$ is the energy of a free hole in the quantum well of
width W in the presence of a magnetic field B directed along the
confinement direction. The dependence of the binding energy of
the positively charged exciton spin-singlet state on the well
width is shown in Fig.~\ref{xp-width} (solid curve) in the
absence of a magnetic field. Our results are also compared with
the Monte Carlo calculations of Ref.~\onlinecite{Japan} (dotted
curve). The results from the two theories differ by about 0.1 meV
at $W=200$~{\AA}. One reason for this difference could be
attributed to the different choice for the value of the static
dielectric constant; we took it equal to 12.58 while in
Ref.~\onlinecite{Japan} it was taken equal to 12.5. Notice that
while the binding energy is decreasing with increasing well width
according to both theories our calculation shows a faster decrease
of the positively charged exciton binding energy with increasing
well width, this behaviour is particularly strong for the $X^+$
system. The symbols in Fig.~\ref{xp-width} represent the
experimentally measured values for the spin-singlet state binding
energy of the $X^+$ for different quantum well width and
different experiments.\cite{Shields95-1,Finkelstein,Osborne}
Notice that these results are in good agreement with the theory of
Ref. \onlinecite{Japan}, and agree with ours in the nominal error
of the theory, i.e. $\pm 0.1 meV$. This error is estimated
considering the digit to which we round off our result for the
total energies, however due to the very good convergence of the
calculation we expect the ``real error'' to be smaller than the
estimated one. In general we would expect that the experimental
energies are larger than the theoretical calculated ones because:
1) a non zero density of holes leads to a non zero Fermi energy
$E_F$ and it has been shown that the experimentally determined
binding energy is in fact $E_F+E_B(X^+,B=0)$; and 2) quantum well
width fluctuations will localize the trion. Both effects lead to a
larger binding energy with respect to the one calculated
theoretically for a free translational invariant trion.

In Fig.~\ref{xp-width}, we also report the dependence of the
negatively charged exciton binding energy on the well width
(dashed curve). Notice that the $X^+$ binding energy is larger
(about $20 \%$) as compared to the $X^-$ binding energy, both for
our theory and for the one of Ref.~\onlinecite{Japan}. Moreover
this result is also in agreement with the experimental results
which showed that the energy of the $X^+$ is
larger\cite{Shields95-2,Shields95-1} or equal\cite{Glasberg} to
the one of $X^-$. St{\'e}b{\'e} and Moradi\cite{Stebe2000} found,
on the other hand, that for a 300~{\AA} wide \ quantum well the
binding energy of $X^+$ is lower than the one of $X^-$ which is
opposite to our conclusion and to those of
Ref.~\onlinecite{Japan}. It should be stressed that the
theoretical value for the trion binding energy depends not only
on the obtained value for the trion energy but also on the
exciton energy. In out theory, the latter are both upper bounds to
the exact result and consequently the trion binding energy is
neither an upper nor a lower bound to the exact value. For
example, an overestimation of the exciton energy will lead to an
overestimation of the trion binding energy.

\section{Magnetic field dependence of the trion properties}
\label{first-PPP}

Next, we compare the dependence of the binding energy of the
positively charged exciton on the magnetic field with the one of
the negatively charged exciton, for a quantum well of width 100
{\AA} \ (see Fig.~\ref{100AA}). Notice that while for $B<1$~T we
obtain $E_B(X^+,B)>E_B(X^-,B)$, which is in agreement with the
$B=0$~T behaviour as shown in Fig.~\ref{xp-width}. For $B>1$~T the
relation between the two energies is reversed and the negatively
charged exciton becomes more strongly bound.  Moreover while the
$X^-$ singlet binding energy quickly increases up to about 2.5
meV for $B=10$ T after which it saturates, the $X^+$ binding
energy increases slowly from about 1.2~meV at $B=0$ up to about
1.5 meV for $B=7$ T, where it starts to decrease slowly. We found
a similar behaviour for the binding energy in the case of a
quantum well of width 200 {\AA} (see Fig.~\ref{200-bin}). For the
latter quantum well width we also calculated the spin-triplet
state with $L=1$ (dashed-dotted line in Fig.~\ref{200-bin}). The
triplet state is unbound at low magnetic field and becomes bound
around 1 T. We also find that the binding energy of this state
increases rather fast with increasing magnetic field and
eventually we observe a crossing around $B=15$ T. This behaviour
is consistent with the one found for the negatively charged
exciton where we found that when the hole mass was much larger
than the electron one no spin-singlet spin-triplet transition was
observed. However, when the hole mass was decreased considerably
such singlet-triplet transition occurred. In that case the hole
mass was always larger than the electron one, however it is
reasonable to assume that decreasing further the hole mass the
singlet-triplet transition would still be observed. Here the
roles of the hole and the electron are switched and the electron
mass is much smaller than the hole one. As a consequence this is
then consistent with the fact that we now find a singlet-triplet
transition.

The average distance between pairs of particles of the trion in
the plane orthogonal to the quantum well axis, i.e.
$<\rho_{ij}^2>^{1/2}$, is shown in Fig~\ref{average-chapter4} as
function of the magnetic field for quantum well of width 200
{\AA}. Notice that for $B=0$ the average distance between the
positively charged exciton pairs, i.e.\ hole-hole and
electron-hole ones, are smaller than the correspondent one for
the negatively charged exciton. In particular the difference is
more pronounced for the pair of particles with the same charge
(i.e.\ hole-hole in case of $X^+$ versus electron-electron in
$X^-$). Since the attractive and the repulsive contribution to the
potential are inversely proportional to the average interparticle
distance, the behaviour at $B=0$ explains the fact that we find a
lower binding energy for the $X^-$ than for the $X^+$ at $B=0$.
Notice that around 1~T the average distance of the electron-hole
pair in $X^-$ crosses the one of $X^+$, this is the same magnetic
field at which the $X^-$ binding energy becomes larger than the
$X^+$ binding energy.  For increasing magnetic field the ratio
between the average distance between the electron-electron and
electron-hole pair in $X^-$ is rather constant. Looking at the
corresponding pairs in $X^+$ we see that the distance between the
electron and hole diminishes with increasing magnetic field but
less than the one between the two holes. This could explain the
slow increase and then the decrease already at low magnetic
fields of the $X^+$ binding energy against the fast increase and
saturation of the $X^-$ binding energy.

\begin{figure}[tbp]
 \caption{The
pair correlation function for the spin singlet state of the
positively charged exciton in a 200 {\AA} \ wide quantum well as
function of the magnetic field.} \label{corr-xp} 
\end{figure}
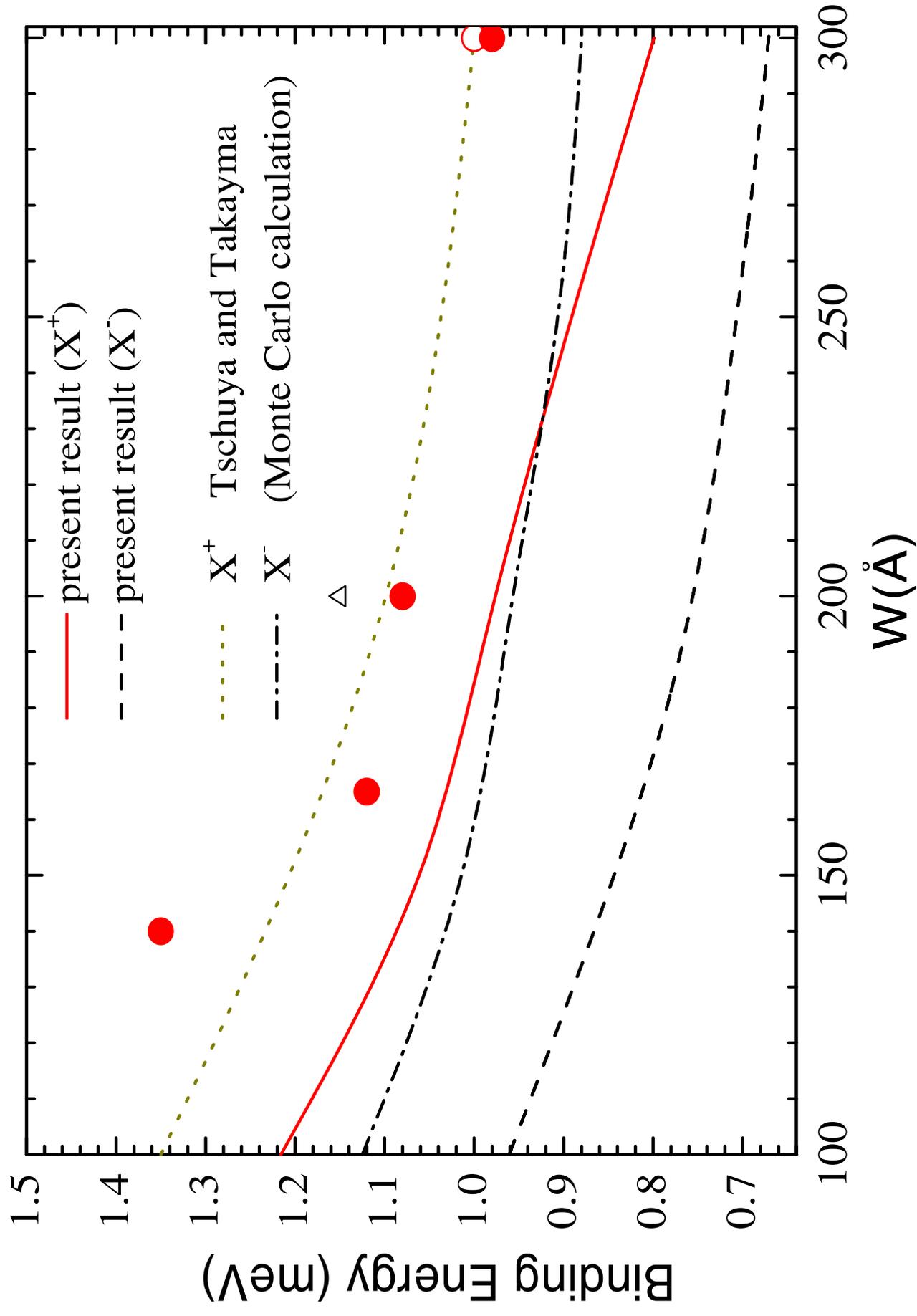

We calculated also the pair correlation function for the positive
trion in the confinement direction,
$g_{ij}(z)=<\delta(z-|z_i-z_j|)>$, and in the orthogonal plane,
$g_{ij}(\rho)=<\delta(\rho-|\vec{\rho}_i-\vec{\rho}_j|)>$, both at
$B=0$ and at $B=4$ T (see Fig.~\ref{corr-xp}). Notice that along
the $z$-axis, Fig.~\ref{corr-xp}(a), both the hole-hole and the
electron-hole pair correlation function have a peak around the
center of the quantum well, which is in agreement with the fact
that the Coulomb interaction plays a minor role as compared to the
confinement due to the presence of the quantum well. Notice also
that, as expected, the presence of a magnetic field has only a
very small influence on the correlation function in the
$z$-direction. In the $\rho$-plane, Fig.~\ref{corr-xp}(b),  we
observe that the electron-hole pair correlation is peaked around
zero, i.e. the electron and the hole maximize their interaction
by staying as close as possible to each other. The hole-hole pair
correlation function has instead a peak around the average pair
inter-particle distance. When a magnetic field is applied the
peak of the hole-hole pair correlation function is shifted towards
lower $\rho$. This is consistent with the fact that the average
pair inter-particle distance decreases with increasing magnetic
field.

\section{Comparison with experiment}
\label{second-P} In Fig.~\ref{200-bin}, we compare the binding
energy of the $X^+$ and of the $X^-$ singlet state for a 200
{\AA} \ wide quantum well with the experimental results obtained
by Glasberg {\it et al.} \cite{Glasberg} and the very recent
results Yusa {\it et al.}\cite{Yusa} (open triangles). When
comparing the experimental results with our theoretical results
we notice that for $B=0$ the experimental and theoretical result
for the negatively charged exciton binding energy differ by about
0.3 meV, which may be due to a non-zero density effect and/or to
localization induced by quantum well width fluctuation. In the
range $1$ T$<B<7$ T there is very good agreement between theory
and experiment. For $B>8$ T the experimental binding energy
saturates, while this occurs at higher fields for our theoretical
results. The agreement of the $X^-$ singlet energy with the very
recent experimental results of the same group\cite{Yusa} is on
the other hand remarkable. Not only theoretical but also
experimental results improve with time.

For the $X^+$ spin-singlet state binding energy, the agreement is
rather good over the whole magnetic field range at which
experimental results are available. However at low magnetic
fields the theoretical binding energy underestimates the
experimental value. It has been argued that this effect can be
attributed to localization effects due to quantum well width
fluctuations. Notice that localization effects are less important
on the positively charged exciton than on the negatively charged
exciton, which is consistent with the larger mass of the holes
and which leads to a less extended trion (see
Fig.~\ref{average-chapter4}). For the $X^+$ spin-triplet state a
comparison with the experimental data by Glasberg {\it et
al.}\cite{Glasberg} shows a good agreement for magnetic fields up
to about 3 T. Beyond 3 T the experimental energy saturates while
the theory still predicts an increasing binding energy with
increasing magnetic field.
%
\section{Conclusion}
\label{third-P} The present work is the first theoretical work in
which a detailed comparison is made between  experimental and
theoretical singlet and triplet binding energies of the positively
charged exciton. The theoretical results explain the different
magnetic field dependence of the $X^-$ and $X^+$ ground state
binding energy, which was observed experimentally. Namely the
$X^+$ singlet has a very weak magnetic field dependence while the
$X^-$ singlet binding energy increases rapidly in the low magnetic
field region and saturates at higher fields. We find good
qualitative and quantitative agreement with the available
experimental results. The $X^+$ triplet binding energy on the
other hand increases with magnetic field and we predict for a
quantum well of width 200 \AA \ a singlet-triplet crossing at
$B\approx 15$ T. The experiments of Ref.~\onlinecite{Glasberg}
show a saturated triplet binding energy for $B> 3$ T, which we do
not find in our theoretical results.
\section{Acknowledgment}

Part of this work is supported by the Flemish Science Foundation
(FWO-Vl), the `Interuniversity Poles of Attraction Program -
Belgian State, Prime Minister's Office - Federal Office for
Scientific, Technical and Cultural Affairs' and the
Flemish-Hungarian cultural exchange program. K. Varga was
supported by the U. S. Department of Energy, Nuclear Physics
Division, under contract No. W-31-109-ENG-39 and OTKA grant No.
T029003 (Hungary). We thank Go Yusa for providing us with his
experimental data for the 200 \AA \ wide quantum well before
publication.

\begin{figure}[tb]
\caption{ Dependence of the binding energy of the positively and
negatively charged exciton on the well width at zero magnetic
field. The symbols represent experimental data for the $X^+$ from
Ref. 18 (full circles), Ref. 14 (open circle) and Ref. 16
(triangle).}\vspace{-0.5cm} \label{xp-width}
\end{figure}

\begin{figure}[tb]
 \caption{The
binding energy of a negatively and a positively charged exciton in
a 100 {\AA} \ wide quantum well as a function of the applied
magnetic field.} \label{100AA}
\end{figure}

\begin{figure}[tb]
 \caption{The
binding energy of a negatively and positively charged exciton in
a 200 {\AA} \ wide quantum well. The open and full circles as
well as the diamonds are the experimental data from Ref. 19, the
triangles are experimental data from Ref. 24. The bars represent
the nominal error of our calculation.}\vspace{-0.5cm}
\label{200-bin}
\end{figure}

\begin{figure}[tb]
 \caption{The
average pair distance in a positively and in a negatively charged
exciton in a 200 {\AA} \ wide quantum well as function of the
magnetic field.} \label{average-chapter4} 
\end{figure}

\end{document}